\documentclass[conference]{IEEEtran}
\IEEEoverridecommandlockouts
\usepackage{cite}
\usepackage{quantikz}
\usepackage{tikz}
\usepackage{amsmath,amssymb,amsfonts}
\usepackage{algorithmic}
\usepackage{graphicx}
\usepackage{textcomp}
\usepackage{xcolor}

\def\BibTeX{{\rm B\kern-.05em{\sc i\kern-.025em b}\kern-.08em
    T\kern-.1667em\lower.7ex\hbox{E}\kern-.125emX}}
\begin{document}

\title{The Questionable Influence of Entanglement in Quantum Optimisation Algorithms\\
}

\author{
\IEEEauthorblockN{%
Tobias Rohe\IEEEauthorrefmark{1},
Daniëlle Schuman\IEEEauthorrefmark{1},
Jonas Nüßlein\IEEEauthorrefmark{1},
Leo Sünkel\IEEEauthorrefmark{1},
Jonas Stein\IEEEauthorrefmark{1}\IEEEauthorrefmark{2}
and Claudia Linnhoff-Popien\IEEEauthorrefmark{1}}
\IEEEauthorblockA{\IEEEauthorrefmark{1}\textit{LMU Munich, Institute for Computer Science, Munich, Germany}}
\IEEEauthorblockA{\IEEEauthorrefmark{2}\textit{Aqarios GmbH, Munich, Germany}}
\IEEEauthorblockA{\{tobias.rohe, danielle.schuman, jonas.nuesslein, leo.suenkel, jonas.stein, linnhoff\}@ifi.lmu.de}
}

\maketitle

\begin{abstract}
The performance of the Variational Quantum Eigensolver (VQE) is promising compared to other quantum algorithms, but also depends significantly on the appropriate design of the underlying quantum circuit. Recent research by Bowles, Ahmend \& Schuld, 2024 \cite{bowles2024better} raises questions about the effectiveness of entanglement in circuits for quantum machine learning algorithms. In our paper we want to address questions about the effectiveness of state preparation via Hadamard gates and entanglement via CNOT gates in the realm of quantum optimisation. We have constructed a total of eight different circuits, varying in implementation details, solving a total of 100 randomly generated MaxCut problems. Our results show no improvement with Hadamard gates applied at the beginning of the circuits. Furthermore, also entanglement shows no positive effect on the solution quality in our small scale experiments. In contrast, the investigated circuits that used entanglement generally showed lower, as well as deteriorating results when the number of circuit layers is increased. Based on our results, we hypothesise that entanglement can play a coordinating role, such that changes in individual parameters are distributed across multiple qubits in quantum circuits, but that this positive effect can quickly be overdosed and turned negative. The verification of this hypothesis represents a challenge for future research and can have a considerable influence on the development of new hybrid algorithms.  
\end{abstract}


\section{Introduction} \label{introduction}
Considerable progress has been made in quantum hardware and quantum algorithms for many years. The search for an optimal architecture \cite{kolle2023disentangling, altmann2023challenges} and efficient Ansatz-design \cite{sim2019expressibility} is proving to be a challenge. In this context, often the terms \textit{superposition} and \textit{entanglement} from quantum mechanics are referenced, which are closely related to the Hadamard gate and the CNOT gate in the field of quantum algorithms and circuits. Since questions regarding their effectiveness have repeatedly arisen in the literature \cite{bowles2024better, diez2021quantum, marrero2021entanglement, patti2021entanglement}, this paper takes a closer look at their direct effect and shows how they influence the convergence behaviour and the solution quality of the variational quantum eigensolver (VQE) in the context of quantum optimisation. 
We apply different circuit designs to randomly generated MaxCut problem instances to separate the effects of the individual circuit elements. This gives us an indication that the use of Hadamard gates for state initialisation at the circuit beginning is not always advantageous. The results also show that the use of entanglement does not automatically lead to better solutions, but actually worsens the solution quality in our case, with its negative impact becoming even stronger as the number of applied circuit layers increases. However, our results also indicate that a possible coordinative effect can be attributed to entanglement, in that changing individual parameters changes the quantum state of several qubits. In connection with this, we hypothesise that, also based on the findings from other literature~\cite{nannicini2019performance, miki2022variational, mcclean2021low, diez2021quantum, chen2022much, sreedhar2022quantum, marrero2021entanglement, patti2021entanglement, sharma2022trainability, bowles2024better, kim2022quantum, kim2022entanglement, woitzik2020entanglement, zhang2023single, gross2009most}, entanglement can easily be overdosed and lead to negative effects. Cautious use is therefore advisable, although according to current research~\cite{kim2022entanglement, woitzik2020entanglement, nannicini2019performance, qian2024information, diez2021quantum, chen2022much}, no clear statement can be made about the correct dose.
With this analysis, we provide a deeper understanding of the use of Hadamard gates and entanglement in hybrid quantum algorithms to promote efficient utilisation and design decisions of quantum resources in the future.

The paper is structured as follows: In the subsequent Sec.~\ref{background}, we will deal with the necessary background for this work. The methodology used is then explained in Sec.~\ref{methodology}. Sec.~\ref{results} follows with the results achieved, while Sec.~\ref{discussion} critically analyses and discusses these and identifies the limitations of our research. Our paper ends with a conclusion in Sec.~\ref{conclusion}.

\section{Background} \label{background}

\subsection{MaxCut Problem}
In graph theory, the Maximum Cut (MaxCut) problem seeks for the partitioning of the vertices $v \in V$ of the graph $G = (V, E)$ into two complementary sets with the objective to maximise the number of edges $\{v, u\} \in E$ between the two sets $S$ and $T$. The number of nodes is denoted as $n$, whereby the numbering of the nodes is defined as $V = \{0,...,n-1\}$. In our study, we address the unweighted MaxCut problem \cite{crescenzi2001weighted} by assigning a uniform weight of $w_{v,u} = 1$ to all edges. Because of the binary decision whether a node is included in the one or the other set, the MaxCut problem is particularly suited for an efficient quantum encoding, node $v$ encoded as $x_v= 0$ for $v \in S$, or $x_v= 1$ for $v \in T$. The cost function to minimise is given as \begin{equation}
C(x) = -\sum_{(u,v) \in E} x_u + x_v - 2x_u x_v.
\end{equation}
The Goemans and Williamson algorithm provides a guaranteed approximation ratio of 0.8786 for this NP-hard combinatorial optimisation problem \cite{goemans1995improved}. The MaxCut problem is characterised by its efficacious quantum encoding, assured classical approximation bounds, and its extensive use for the evaluation of quantum algorithms.
%It is characterised by its efficacious quantum encoding, assured classical approximation bounds, and its prevalent application in the evaluation of quantum algorithms.

\subsection{Variational Quantum Eigensolver}
The VQE was proposed by Peruzzo et al. in 2014 \cite{peruzzo2014variational} and belongs to the category of hybrid quantum algorithms. The algorithm is composed of a quantum part and a classical part, which are both executed in a loop. The quantum part is formed by a parameterised quantum circuit (PQC) $U\left(\theta\right)$, also called Ansatz, which prepares the quantum state $\ket{\psi\left(\theta\right)}\coloneqq U\left(\theta\right)\ket{0}^{\otimes n}$. The quantum circuit described here acts on $n$ qubits and is dependent on the gate parameters $\theta$. These gate parameters $\theta$ are being optimised in the classical part of the algorithm, which closes the quantum-classical loop. The optimisation of the parameters aims to minimise the cost function of the underlying problem, which is programmed and given as Hamiltonian $\hat{H}_C = \sum_{(u,v) \in E}   \sigma_u^z \sigma_v^z$ to the algorithm. The state resulting from the parameter optimisation finally approximates the ground state of the given Hamiltonian $\hat{H}_C$, which should then correspond to the global minima of the cost function. 

The VQE algorithm is particularly suited for chemical problems, like estimating the ground state of a Hamiltonian~\cite{moll2018quantum}, but can also be utilised in the areas of logistics, supply chain, finance and many more~\cite{cerezo2022variational, tilly2022variational, kandala2017hardware, wang2019accelerated, griffin2021quantum}.

\subsection{Related Work}

Related work~\cite{nannicini2019performance, miki2022variational, mcclean2021low, diez2021quantum, chen2022much, sreedhar2022quantum, marrero2021entanglement, patti2021entanglement, sharma2022trainability, bowles2024better, kim2022quantum, kim2022entanglement, woitzik2020entanglement, zhang2023single, gross2009most} finds that increasing the amount of entanglement is not always advantageous when solving problems with variational quantum approaches: Already in 2009, Gross et al.~\cite{gross2009most} proved that the majority of all possible quantum states is ``too entangled'' to be useful for quantum computing in general, given one has the goal to obtain a practical advantage over classical computing. Recent literature furthermore shows that while entanglement in the earlier layers of a variational circuit can be beneficial, as it increases the expressibility of the circuit and the non-locality of information in it, there seems to be a number of entangling layers $l_s$ at which the improvement in solution accuracy or convergence speed that can be gained by increasing entanglement ``saturates''~\cite{sim2019expressibility, diez2021quantum, nakhl2024calibrating, kim2022entanglement, dupont2022entanglement}. Using more than $l_s$ entangling layers will decrease the convergence speed and / or solution quality of the algorithm~\cite{kim2022entanglement}. This is due to the solution landscape of the problem becoming less curved with increasing amounts of entanglement, leading to the formation of barren plateaus in the extreme case~\cite{kim2022quantum, kim2022entanglement, marrero2021entanglement, patti2021entanglement, sharma2022trainability, bowles2024better, wiersema2020exploring}. The right number of entangling layers $l_s$ seems to depend mainly on the structure of the problem to be solved, specifically the separability of the ground state of the Hamiltonian describing the problem as well as the Hamiltonian's density~\cite{woitzik2020entanglement, kim2022entanglement, qian2024information, diez2021quantum, chen2022much}. In case of many binary optimisation problems, this Hamiltonian is a diagonal one with a classical ground state~\cite{teramoto2023role}. Thus, when solving sparse problem instances of this type with standard QAOA or other generic VQE approaches, it can be advantageous in practice to have low entanglement, or even none at all, towards the end of the circuit~\cite{teramoto2023role, kim2022quantum, zhang2023single, nannicini2019performance, chen2022much, miki2022variational, sreedhar2022quantum, diez2021quantum, qian2024information}. Especially for the (weighted) MaxCut problem, while related work on QAOA approaches finds some entanglement in the early circuit layers to be beneficial for solving both problem variants, work using generic VQE approaches to solve the weighted MaxCut problem does not find any added value in using entanglement at all in their experiments~\cite{sreedhar2022quantum, chen2022much, nannicini2019performance, miki2022variational}. Another potentially influential factor on the formation of entanglement, however, seems to be the classical optimiser used~\cite{woitzik2020entanglement, kim2022entanglement}. Here, differences between e.g. gradient-based and gradient-free optimisers remain to be investigated~\cite{kim2022entanglement}. Given the above-mentioned stark differences between different problem formulations and VQE approaches regarding the influence of entanglement, this work contributes an experimental case study on solving the unweighted MaxCut problem with generic VQE approaches, using different numbers of entangling layers. Furthermore, we specifically investigate the effect of a state initialisation using Hadamard gates on the different VQE circuits' performances.

\section{Methodology} \label{methodology}
\begin{figure*}[!ht]
\centering % This will center the figure content
% Circuit a: RY-Circuit
\begin{minipage}[b]{0.48\linewidth} % [b] aligns at the bottom
\centering
\begin{quantikz}[row sep={0.7cm,between origins}, column sep=0.35cm]
\lstick{\ket{0}} & \gate{H}\gategroup[4,steps=1,style={dashed, inner xsep=1pt,fill=blue!20},background,label style={label position=above,yshift=+0.3cm}]{{\textcircled{\raisebox{-0.9pt}{1}}}} & \qw & \gate{R_Y(\theta_0)}\gategroup[4,steps=1,style={dashed, inner xsep=1pt,fill=green!20},background,label style={label position=above,yshift=+0.3cm}]{{\textcircled{\raisebox{-0.9pt}{2}}}} & \qw & \meter{}\gategroup[4,steps=1,style={dashed, inner xsep=1pt,fill=red!20},background,label style={label position=above,yshift=+0.3cm}]{{\textcircled{\raisebox{-0.9pt}{3}}}} & \qw \\
\lstick{\ket{0}} & \gate{H} & \qw & \gate{R_Y(\theta_1)} & \qw & \meter{} & \qw \\
\lstick{\ket{0}} & \gate{H} & \qw & \gate{R_Y(\theta_2)} & \qw & \meter{} & \qw \\
\lstick{\ket{0}} & \gate{H} & \qw & \gate{R_Y(\theta_3)} & \qw & \meter{} & \qw
\end{quantikz}
\\[1ex]
\textbf{a. (H)RY-Circuit}
\end{minipage}\hfill % This percent sign is crucial to prevent breaking into a new line
% Circuit b: RYCNOT-Circuit
\begin{minipage}[b]{0.48\linewidth}
\centering
\begin{quantikz}[row sep={0.7cm,between origins}, column sep=0.35cm]
\lstick{\ket{0}} & \gate{H}\gategroup[4,steps=1,style={dashed, inner xsep=1pt,fill=blue!20},background,label style={label position=above,yshift=+0.3cm}]{{\textcircled{\raisebox{-0.9pt}{1}}}} & \qw & \gate{R_Y(\theta_0)}\gategroup[4,steps=5,style={dashed, inner xsep=1pt,fill=green!20},background,label style={label position=above,yshift=+0.3cm}]{{\textcircled{\raisebox{-0.9pt}{2}}}} & \ctrl{1} & \qw & \qw & \targ{} & \qw & \meter{}\gategroup[4,steps=1,style={dashed, inner xsep=1pt,fill=red!20},background,label style={label position=above,yshift=+0.3cm}]{{\textcircled{\raisebox{-0.9pt}{3}}}} & \qw \\
\lstick{\ket{0}} & \gate{H} & \qw & \gate{R_Y(\theta_1)} & \targ{} & \ctrl{1} & \qw & \qw & \qw & \meter{} & \qw \\
\lstick{\ket{0}} & \gate{H} & \qw & \gate{R_Y(\theta_2)} & \qw & \targ{} & \ctrl{1} & \qw & \qw & \meter{} & \qw \\
\lstick{\ket{0}} & \gate{H} & \qw & \gate{R_Y(\theta_3)} & \qw & \qw & \targ{} & \ctrl{-3} & \qw & \meter{} & \qw
\end{quantikz}
\\[1ex]
\textbf{b. (H)RYCNOT-Circuit}
\end{minipage}

% Ensure there's a clear separation before starting the next row of circuits
\vspace{0.6cm}
% Circuit c: RYRX-Circuit
\begin{minipage}[b]{0.48\linewidth}
\centering
\begin{quantikz}[row sep={0.7cm,between origins}, column sep=0.25cm]
\lstick{\ket{0}} & \gate{H}\gategroup[4,steps=1,style={dashed, fill=blue!20, inner xsep=1pt},background,label style={label position=above,yshift=+0.3cm}]{{\textcircled{\raisebox{-0.9pt}{1}}}} & \qw & \gate{R_Y(\theta_0)}\gategroup[4,steps=2,style={dashed, fill=green!20, inner xsep=1pt},background,label style={label position=above,yshift=+0.3cm}]{{\textcircled{\raisebox{-0.9pt}{2}}}} & \gate{R_X(\theta_4)} & \qw & \meter{}\gategroup[4,steps=1,style={dashed, fill=red!20, inner xsep=1pt},background,label style={label position=above,yshift=+0.3cm}]{{\textcircled{\raisebox{-0.9pt}{3}}}} & \qw \\
\lstick{\ket{0}} & \gate{H} & \qw & \gate{R_Y(\theta_1)} & \gate{R_X(\theta_5)} & \qw & \meter{} & \qw \\
\lstick{\ket{0}} & \gate{H} & \qw & \gate{R_Y(\theta_2)} & \gate{R_X(\theta_6)} & \qw & \meter{} & \qw \\
\lstick{\ket{0}} & \gate{H} & \qw & \gate{R_Y(\theta_3)} & \gate{R_X(\theta_7)} & \qw & \meter{} & \qw
\end{quantikz}
\\[1ex]
\textbf{c. (H)RYRX-Circuit}
\end{minipage}\hfill
% Circuit d: RYRZCNOT-Circuit
\begin{minipage}[b]{0.48\textwidth}
\centering
\begin{quantikz}[row sep={0.7cm,between origins}, column sep=0.20cm]
\lstick{\ket{0}} & \gate{H}\gategroup[4,steps=1,style={dashed, fill=blue!20, inner xsep=1pt},background,label style={label position=above,yshift=+0.3cm}]{{\textcircled{\raisebox{-0.9pt}{1}}}} & \qw & \gate{R_Y(\theta_0)}\gategroup[4,steps=6,style={dashed, fill=green!20, inner xsep=1pt},background,label style={label position=above,yshift=+0.3cm}]{{\textcircled{\raisebox{-0.9pt}{2}}}} & \gate{R_X(\theta_4)} & \ctrl{1} & \qw & \qw & \targ{} & \qw & \meter{}\gategroup[4,steps=1,style={dashed, fill=red!20, inner xsep=1pt},background,label style={label position=above,yshift=+0.3cm}]{{\textcircled{\raisebox{-0.9pt}{3}}}} & \qw \\
\lstick{\ket{0}} & \gate{H} & \qw & \gate{R_Y(\theta_1)} & \gate{R_X(\theta_5)} & \targ{} & \ctrl{1} & \qw & \qw & \qw & \meter{} & \qw \\
\lstick{\ket{0}} & \gate{H} & \qw & \gate{R_Y(\theta_2)} & \gate{R_X(\theta_6)} & \qw & \targ{} & \ctrl{1} & \qw & \qw & \meter{} & \qw \\
\lstick{\ket{0}} & \gate{H} & \qw & \gate{R_Y(\theta_3)} & \gate{R_X(\theta_7)} & \qw & \qw & \targ{} & \ctrl{-3} & \qw & \meter{} & \qw
\end{quantikz}
\\[1ex]
\textbf{d. (H)RYRXCNOT-Circuit}
\end{minipage}
\caption{Description of the applied quantum circuits. We employ four different quantum circuits, a. (H)RY-Circuit, b. (H)RYCNOT-Circuit, c. (H)RYRX-Circuit, d. (H)RYRXCNOT-Circuit. Each circuit is composed of three sections: \textcircled{1} an optional ladder of Hadamard gates, \textcircled{2} one or several characteristic layers of the circuit, of which only one layer is shown here, \textcircled{3} the applied measurement scheme. Depending on the implementation of the Hadamard-ladder, the circuits' name begins with the letter "H".} % Add your descriptive caption here
\label{fig:quantum_circuits} % Optional: Adds a label for referencing the figure
\end{figure*}
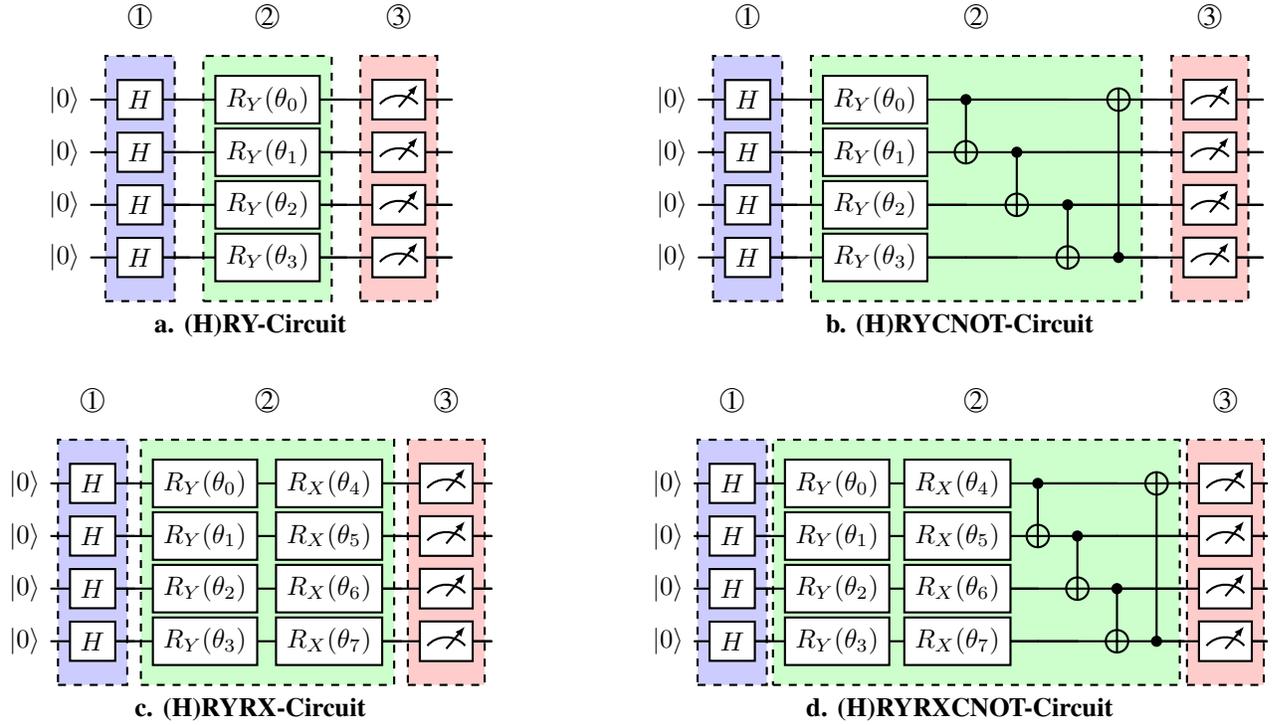
 
In this study, a total of $100$ connected Erdős–Rényi graphs \cite{erdHos1960evolution} were generated, each featuring $10$ nodes with an edge probability set at $p=0.4$. We then solved the MaxCut problem on these randomly generated graphs. To derive the optimal solutions of each problem instance we conducted a brute-force search.\\
As the cornerstone of our research, the VQE algorithm was implemented, utilising $10$ qubits to facilitate a direct mapping of each graph node to a corresponding qubit. In Fig. \ref{fig:quantum_circuits}, all utilised parameterised quantum circuits are depicted for a four qubit implementation. The implementation and structure for the 10 qubit case can be easily derived from generalising their architecture. Each circuit tested consists of three different sections. The first section \textcircled{1}, highlighted in blue, consists of Hadamard gates, which are applied to each qubit individually. This section has an optional character as we also define a sub-variant of the base-circuit employed without this ladder of Hadamards. An ``H'' prefix in circuit names denotes the inclusion of the Hadamard gate ladder, marking the circuits that utilise this sequence. The second section of each circuit \textcircled{2}, coloured in green, shows a single layer that is unique to the four different base-circuits employed. We have chosen the middle layers so that they are as diverse as possible and cover the most varied aspects of circuit design, but without testing too many circuits. The circuits vary in terms of entanglement (\textit{a.} and \textit{c.} versus \textit{b.} and \textit{d.}), but also in terms of the number of different single rotational gates applied (\textit{a.} and \textit{b.} versus \textit{c.} and \textit{d.}). Given that our circuits are constructed in a layered approach, the second section is constructed, depending on the layer-parameter $l \in \mathbb{N}$, by repeatedly applying the layer illustrated here $l$ times. Therefore, a circuit described as having, for example, five layers ($l=5$) implies that this unique \textcircled{2} section is repeated five times in the underlying circuit. The final section \textcircled{3}, coloured in red, entails a layer of measurement operations applied to each qubit, measuring the prepared quantum state. This third section is applied in the same way in all circuits.\\
The four examined circuits of our study are differentiated by their second section, while their sub-variants also vary through the application or omission of the first section. Circuits \textit{a.} and \textit{b.} exclusively utilise rotational Y-Gates in their second section, with circuit \textit{b.} additionally incorporating a circular CNOT entanglement. Conversely, circuits \textit{c.} and \textit{d.} employ both rotational Y and X gates, with circuit \textit{d.} applying a circular CNOT entanglement. The design of the four circuits was intended to encompass a broad spectrum of configurations, from varying numbers of parameterised gates to the presence or absence of entanglement.\\
Throughout our experiments, we executed varying numbers of layers, specifically $1$, $3$, $5$, $7$, $9$, and $20$ layers, enabling us to draw conclusions regarding the performance of deeper circuit configurations. We used the gradient-free COBYLA optimiser \cite{powell1994direct} for our study, conducting all experiments through classical simulations on the Atos Quantum Learning Machine \cite{myqlmWelcomePage}. For scientific rigour, 10 consecutive seeds ($30$ to $39$) were utilised in all experiments. The subsequently employed \textit{approximation ratio} is defined as the cost of the partition obtained from the VQE execution, divided by the cost of the best solution, found via brute-force search. In this context, since our objective is to maximise the costs, an approximation ratio of $1$ indicates that the optimal solution was found.

\section{Results} \label{results}
 In the subsequent section, we will examine the results of our work, first looking at the convergence behaviour of the algorithm and then focusing on the solution quality achieved.
 
\subsection{Convergence Behaviour}

\begin{figure*}[htb]
  \centering
  \includegraphics[width=\textwidth]{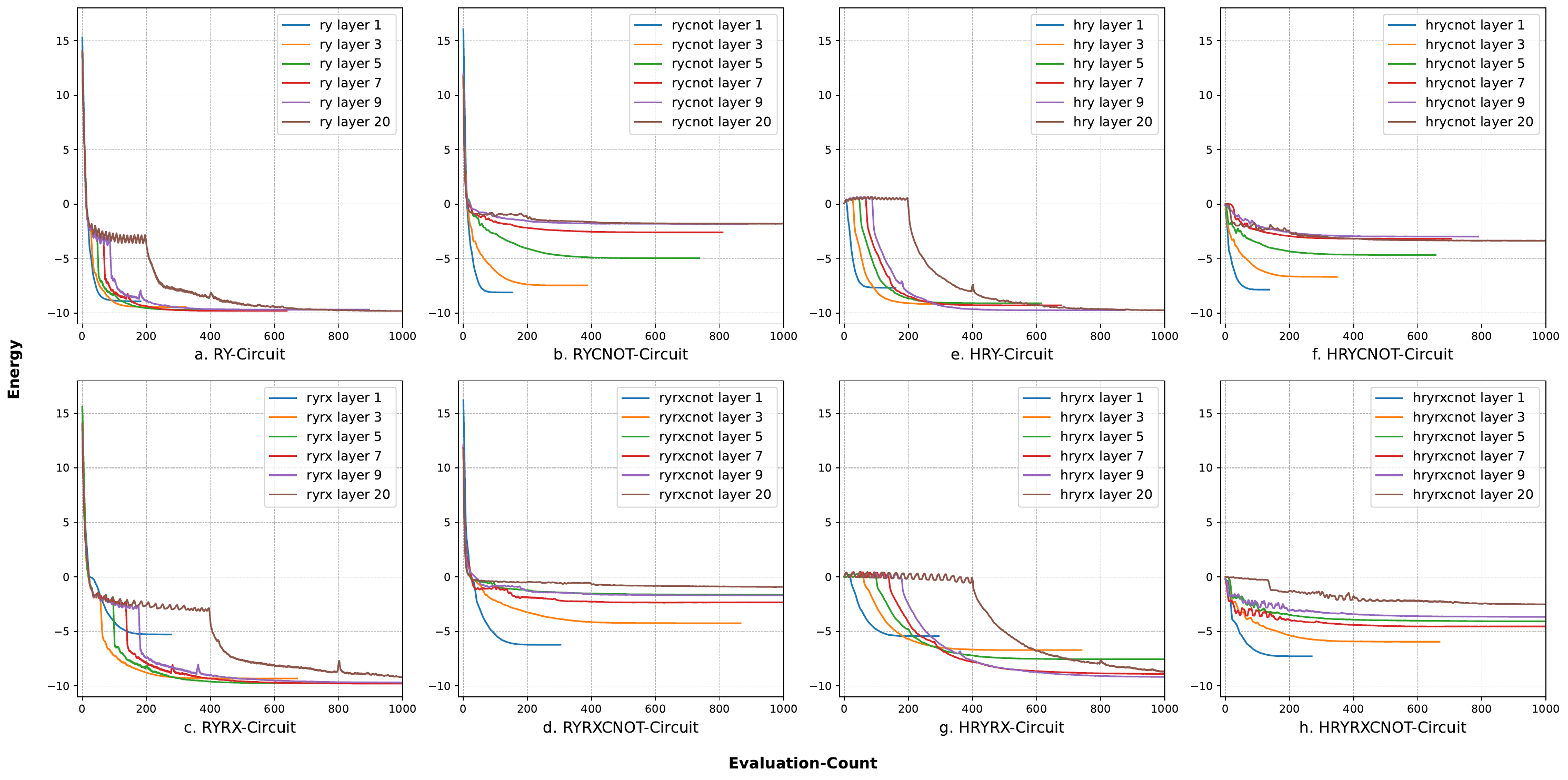}
  \caption{Energy Convergence Chart of VQE Executions. The graph shown here illustrates the convergence behaviour of the various executed VQE instances. Each plot shows a circuit, for which the number of repeating layers varies within the plots. Smaller values equals better.}
  \label{fig:all_convergences}
\end{figure*}

\subsubsection{Utilisation of Hadamard Gates}

Examining the convergence patterns depicted in Fig. \ref{fig:all_convergences}, our initial analysis focuses on the effects of Hadamard gates at the beginning of the circuit, thereby contrasting the circuits labeled \textit{a.}, \textit{b.}, \textit{c.}, and \textit{d.}, with those designated as \textit{e.}, \textit{f.}, \textit{g.}, and \textit{h}. For all circuits using Hadamard gates, we consistently observe a starting value of the convergence series at a level around zero, which is significantly lower, by about $15$ units, than for the circuits without Hadamard layer. A plateau-like phenomenon is observed in circuits that use Hadamard gates but have no form of entanglement in their configurations, see circuits \textit{e.} and \textit{g.}. This particular form of plateau is not evident in the remaining circuits, aside from a possible exception for circuit \textit{h.} at $l=20$. We observe a kind of plateau-like behaviour for non Hadamard enhanced circuits \textit{a.} and \textit{c.} with higher number of layers. In general, it is observed that the deeper the circuit, the more pronounced the plateau becomes. With regards to the convergence level, we do not observe major systematic differences. 

\subsubsection{Utilisation of Entanglement}
Continuing with our examination of Fig. \ref{fig:all_convergences}, we compare the convergence behaviour between circuits that implement entanglement, circuits \textit{b.}, \textit{d.}, \textit{f.}, and \textit{h.}, and those devoid of any entangling operations, circuits \textit{a.}, \textit{c.}, \textit{e.}, and \textit{g}. Generally, no difference is observed regarding the starting level of the convergence series (\texttt{Evaluation Count = $1$}) between the circuits. Nonetheless, variations in the energy levels of convergence are observed. Circuits that incorporate entanglement consistently exhibit poorer performance, as indicated by generally higher energy levels at the time of convergence. Exceptions are seen only in the case of very shallow circuits, where the convergence levels are relatively similar. Additionally, an inverse correlation between the number of layers and circuit efficacy is observed for circuits employing entanglement. 
In circuits without entanglement, so-called product-state Ansätze~\cite{diez2021quantum}, the theoretical premise \cite{kim2021universal} that deeper circuits exhibit improving convergence levels (therefore lower convergence levels), ignoring any noise-effects, is substantiated by our observations. Aside from convergence levels, as to be expected, it is universally observed that the number of layers positively correlate with time to convergence for both entangled and non-entangled circuits.

\subsection{Solution Quality}

\begin{figure*}[htb]
  \centering
  \includegraphics[width=\textwidth]{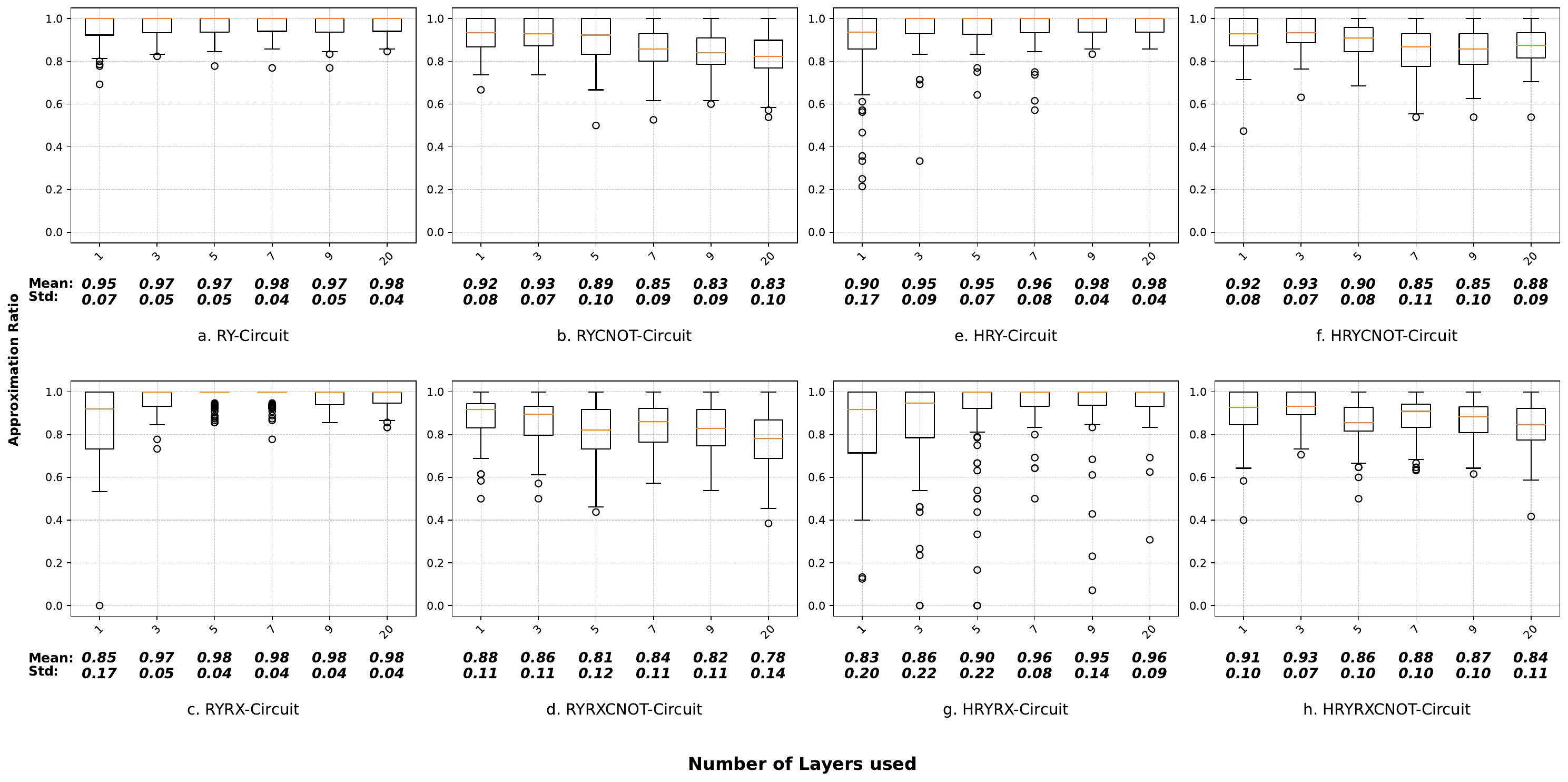}
  \caption{Approximation Ratios achieved by the VQE Executions. The graph shown here visualises the approximation ratios achieved by the various implemented circuits. Higher values equal better results. The respective mean value and standard deviation are given below each box.}
  \label{fig:all_boxplots}
\end{figure*}

In addition to the convergence behaviour of the circuits, we also analyse the achieved solution quality, presented in Fig. \ref{fig:all_boxplots}. Similar to before, we first examine the results with regards to the effect of Hadamard gates in front of the circuit, followed by the examination of the influence of entanglement. 

\subsubsection{Utilisation of Hadamard Gates}
Examining the non-entangled circuits \textit{a.}, \textit{c.}, \textit{e.}, and \textit{g.}, we observe generally slightly superior approximation ratios, equivalent to a better solution quality, for circuits that do not employ a Hadamard gate upfront. For circuits incorporating an entanglement scheme, the trend generally inverts, as incorporating Hadamard gates marginally improves outcomes. For the sake of scientific rigour one has to acknowledge the minimal magnitude of these differences, warranting cautious interpretation here. In terms of the standard deviation of approximation ratios achieved, our observations do not reveal any consistent patterns of difference. Only circuit \textit{g.} exhibits a notably higher standard deviation, especially in scenarios of low circuit depth.

\subsubsection{Utilisation of Entanglement}
Overall, circuits that incorporate entanglement exhibit inferior performance, characterised by lower approximation ratios and greater standard deviations. This disparity becomes particularly pronounced in circuits of greater depth, whereas solution quality for lower depths perform somewhat similar to non-entangled circuits. Examining the progression of solution quality relative to circuit depth, we discern two distinct patterns, clearly distinguishable between circuits with and without entanglement. For circuits devoid of entanglement, an enhancement in performance with increased circuit depth is observed, aligning with our expectations based on results from the literature \cite{kim2021universal}. However, this pattern of enhancement is not evident in circuits featuring entanglement. In these instances, a decline in approximation ratios becomes apparent, as clearly illustrated by the provided data and plots shown in Fig. \ref{fig:all_boxplots}, see circuit \textit{b.}, \textit{d}, \textit{f}, and \textit{h}. Additionally, we do not observe any indication of an U-shaped pattern here, which would indicate that the approximation ratio improves again with sufficient depth.

\section{Discussion \& Limitations} \label{discussion}
In the following, we will first discuss the state initialisation using Hadamard gates and afterwards analyse the influence of entanglement in detail.
 
\subsection{State Initialisation via Hadamard Gates}

Employing a preliminary layer of Hadamard gates can be interpreted as an initialisation technique, initialising the algorithm as if each qubit, hence each node, is simultaneously included in both sets. Consequently, this also explains the lower initial point of convergence observed. 

Mathematically, this is a direct consequence of
\begin{align}
    &&\bra{0}^{\otimes n}\hat{H}_C \ket{0}^{\otimes n}&=|E|,\textnormal{and}\\
    &&\bra{+}^{\otimes n}\hat{H}_C \ket{+}^{\otimes n}&=0,&
\end{align}
as $\hat{H}_C$ does not directly correspond to the number of edges being cut $c$, but rather to $2c-|E|$ (for details, see~\cite{barahona1988application}), and as the expectation value of a random assignment of nodes to partitions yields $c=|E|/2$ (see \cite[Thm. 6.3]{upfal2005probability}).

When analysing the convergence behaviour of the various implementations, we come to the conclusion that a preliminary application of Hadamard gates at the circuit's beginning does not systematically enhance its performance. In certain instances, utilising Hadamard gates initially appears beneficial; however, generally, it requires a case-by-case evaluation and cannot be universally endorsed.

\subsection{Impact of Entanglement}
Through the chosen study design, we have paired each selected circuit with and without entanglement, enabling us to conduct a direct comparison and analyse the effect of entanglement more thoroughly. Generally, it is observed that entangled circuits tend to perform poorer, with only the first layer showing comparable results. We explain this by arguing that in these circuits with low depth, the entanglement does not yet come into play and has no significant influence, which is confirmed by the observation.

As the circuit depth increases, through adding more layers, the performance of entangled versions deteriorates in two respects: First, relative to non-entangled circuits, and second, compared to its less shallow versions. Although there is a short-term improvement in performance from one to three layers in some instances, this benefit is not sustained, and performance quickly deteriorates with further increases in layer count. We attribute this development to the effects of increasing entanglement and the increase in the number of parameters. We hypothesise that the initially beneficial effect of increasing the number of gate parameters is eventually offset by the progressively adverse impact of entanglement in deeper circuits. This is supported by the observation that for non-entangled circuits, which are identical except for the entanglement, the solution quality generally improves with increasing depth and therefore the associated number of parameters.

The question now emerges as to why entangled circuits generally yield poorer, particularly worsening results. A potential, yet speculative, explanation might be that the level of entanglement within the circuit is excessively high, preventing the classical optimiser from effectively steering towards the global optimum. We refer to this phenomenon as “over-entanglement".

A noteworthy observation in this context are the initial plateau's identified at the starting point of the convergence path in Fig. \ref{fig:all_convergences} \textit{a}, \textit{c}, \textit{e}, and \textit{g}. A feature shared by the referenced circuits is the absence of entanglement. Conversely, circuits that incorporate entanglement do not display this plateau. We interpret the plateau as the duration required by the classical optimiser to ascertain the optimal direction for parameter optimisation, as a strong movement towards a significantly better solution can be seen after the plateau has ended. In other words, the energy level does not drop at the flat points (the plateaus), meaning that no significantly better solutions are found in this phase of the optimisation, although the optimisation continues and parameters get changed. If this plateau is especially flat and long, we argue that the classical optimiser had particular problems finding the right optimisation direction. The end of the plateau and the then rapidly decreasing energy level can be interpreted in such a way that a direction to a significantly better optimum has been found and is now being pursued by the optimiser.

Focusing on the initiation point of convergence for circuits devoid of entanglement and Hadamard gates, circuit \textit{a.} and \textit{c.}, the initial optimisation steps to enhance outcomes are readily achievable, as any parameter modification directly yields improved results. However, at a critical juncture, this task becomes more challenging, coinciding with the emergence of the plateau. In circuits excluding entanglement but including Hadamard gates, \textit{e.} and \textit{g.}, this particular critical moment occurs immediately. The implementation of Hadamard gates, as previously mentioned, places the nodes in a state of quasi-membership in both groups simultaneously; yet, through measurement and the inherent probabilistic nature of qubits, a kind of random assignment to the groups is determined. Consequently, optimising the gate parameters proves to be significantly more challenging compared to the initial phase absent of Hadamard gates, where an improvement can quickly be found in the comparatively much worse starting point. It is precisely at this juncture that the optimiser requires time to determine the most effective direction for optimisation.

However, these observations do not fully account for the varied convergence behaviour observed in circuits incorporating entanglement. We speculate that the observed discrepancies originate from a redistributed responsibility for the coordination of the assignment of the nodes of the problem to a certain set. By coordination we mean taking into account the interactions between the nodes within the MaxCut problem. 
In scenarios involving entanglement, the circuit — or more accurately, the entangled quantum state — assumes a critical role in the coordination effort. As the qubits / nodes are entangled with each other, the modification of a single rotating gate parameter however impacts the grouping of several nodes simultaneously, therefore coordinating the assignment of the individual nodes to the two different sets. If, on the other hand, non-entangled circuits are considered, the existence and duration of the plateau can be interpreted as the result of this coordination effort now performed by the classical optimiser. For circuits lacking entanglement, a complex reassignment of several nodes to different groups simultaneously can only be achieved through coordination executed within the classical optimiser.

This rationale would be particularly compelling and relevant, highlighting that entanglement plays a pivotal role in coordinating the optimisation process. Therefore, although entanglement does not present an immediate benefit in the context of our study, we infer that entanglement becomes advantageous and practical for applications demanding extensive coordination, which poses significant challenges for the classical optimiser. While we maintain confidence in our interpretation, further evaluation must determine its validity.

\section{Conclusion} \label{conclusion}
In our study we have investigated the impact of entanglement as well as state initialisation through Hadamard gates on the VQE algorithm and its performance. We have not found any general improvements in solution quality for the MaxCut problem through applying Hadamard gates in front of VQE circuits. However, we observe a negative impact of entanglement on convergence behaviour as well as on solution quality. This negative effect is intensified by increasing the number of layers in the circuit. We explain this with a kind of “over-entanglement” or saturation effect in the degree of entanglement, which was also observed in previous literature. 
Our results provide an experimental basis for the stated theories and help scientists and practitioners to select the appropriate circuit elements. Further, our case study offers a basis for future research regarding the right level of entanglement for selected problems. We believe that the potentially coordinative nature of entanglement requires further investigation to prove or reject this hypothesis and the hypothesis of ``over-entanglement''. The inclusion of additional problems and specifically difficult problem instances in the field of optimisation could provide more insights about the role of entanglement here. The question of whether and how entanglement, often described as ``quantumness'', has a positive effect on the results of quantum machine learning and optimisation is of essential importance for the entire field.

\section*{Acknowledgement}
This paper was partially funded by the German Federal Ministry of Education and Research through the funding program “quantum technologies -- from basic research to market” (contract number: 13N16196). J.S. acknowledges support from the German Federal Ministry for Economic Affairs and Climate Action through the funding program “Quantum Computing -- Applications for the industry” based on the allowance “Development of digital technologies” (contract number: 01MQ22008A).


\vspace{12pt}

\end{document}